\documentclass[%
 reprint,
nobibnotes,
 amsmath,amssymb,twocolumn,
prl,
floatfix,
]{revtex4-1}
\usepackage{multirow}
\usepackage{array}
\newcolumntype{P}[1]{>{\centering\arraybackslash}p{#1}}

\usepackage{amsmath}
\usepackage{amssymb}
\usepackage[capitalize]{cleveref}
\usepackage{graphicx}
\usepackage{dcolumn}
\usepackage{bm}
\usepackage{nomencl}
\makenomenclature
\usepackage{xcolor}
\usepackage{rotfloat}
\usepackage{lipsum}
\floatstyle{plaintop}
\restylefloat{table}

\newcommand{\norm}[1]{\left\lVert#1\right\rVert}

\begin{document}

\title{Discovering Quantum Phase Transitions with Fermionic Neural Networks}

\author{Gino~Cassella$^1$}
  \email{g.cassella20@imperial.ac.uk}
\author{Halvard~Sutterud$^1$}
\author{Sam~Azadi$^4$}
\author{N.D.~Drummond$^3$}
\author{David~Pfau$^{2,1}$}
\author{James~S.~Spencer$^2$}
\author{W.M.C.~Foulkes$^1$}%

\affiliation{%
$^1$Dept.~of Physics, Imperial College London, London SW7 2AZ, United Kingdom}
\affiliation{
 $^2$DeepMind, London N1C 4DJ, United Kingdom
}
\affiliation{
$^3$Dept.~of Physics, Lancaster University, Lancaster LA1 4YB, United Kingdom}
\affiliation{
$^4$Dept.~of Physics, University of Oxford, Oxford OX1 3PU, United Kingdom}

\date{\today}

\begin{abstract}
Deep neural networks have been very successful as highly accurate wave function ans\"atze for variational Monte Carlo calculations of molecular ground states. We present an extension of one such ansatz, FermiNet, to calculations of the ground states of periodic Hamiltonians, and study the homogeneous electron gas. FermiNet calculations of the ground-state energies of small electron gas systems are in excellent agreement with previous initiator full configuration interaction quantum Monte Carlo and diffusion Monte Carlo calculations. We investigate the spin-polarized homogeneous electron gas and demonstrate that the same neural network architecture is capable of accurately representing both the delocalized Fermi liquid state and the localized Wigner crystal state. The network converges on the translationally invariant ground state at high density and spontaneously breaks the symmetry to produce the crystalline ground state at low density, despite being given no \emph{a priori} knowledge that a phase transition exists.
\end{abstract}

\maketitle

The correlated motion of electrons in condensed matter gives rise to rich emergent phenomena. Although these are governed by fundamental quantum mechanical principles known for almost a century, they remain difficult to understand and even harder to predict theoretically or computationally. One of the major themes of modern condensed matter physics is the study of phase transitions caused by electron correlation.

The difficulty of solving the Schr\"odinger equation scales exponentially with particle number in general, so exact solutions for interacting many-electron systems are rarely accessible. This explains why approximate numerical techniques have become such vital tools in the search for exotic zero-temperature phases, providing accurate predictions of experimentally observable quantities in phases already understood qualitatively. Most computational approaches, however, encode prior assumptions about the appropriate phase, which poses a substantial difficulty in predicting previously unknown electronic states. Changes in symmetry or topology are rarely discovered computationally before they have been seen experimentally or proposed on theoretical grounds.

In this Letter, we introduce a neural-network-based approach to predicting the qualitative nature of electronic ground states in condensed matter. We utilize a representation of the wave function, the fermionic neural network (FermiNet) \cite{pfau_ab_2020}, which is capable of representing \emph{any} antisymmetric state \cite{hutter2020representing}, and requires no \textit{a priori} knowledge of the system being studied. Guided by the quantum mechanical variational principle alone, without reference to experimental data, the FermiNet can learn the ground state of a many-body interacting Hamiltonian. Phase transitions are seen by studying changes in the ground state as the parameters of the system are varied.

A significant body of recent work has used machine learning to detect phase transitions in simulated classical \cite{wang2016discovering, carrasquilla2017machine, van2017learning} and quantum \cite{arai2018deep, venderley_machine_2018, zhang_machine_2019} systems, but these studies required a source of external data, looking for patterns characteristic of different phases. Our approach requires only the Hamiltonian. There has also been work using neural network ans\"atze to study lattice models and spin systems, including their phase transitions \cite{carleo_solving_2017, saito2017solving, luo2019backflow, stokes2020phases, astrakhantsev2021broken}, but for applications to many real systems, the wave function must be treated, as in the present work, in continuous space.

The flexibility of the FermiNet hinges on the universal approximation property of neural networks \cite{cybenko_approximation_1989, hornik_approximation_1991}, which makes them a versatile tool for approximating high-dimensional functions and has led to radical advances in many computational fields \cite{krizhevsky2012imagenet, vaswani2017attention, silver2016mastering, jumper2021highly}. This success has motivated the application of neural networks to solving problems across the physical sciences, including quantum mechanics \cite{torlai_neural-network_2018, melko_restricted_2019, schutt2020machine, carleo_solving_2017}. Several neural-network-based wave functions in both first-quantized \cite{pfau_ab_2020, spencer_better_2020, hermann_deep-neural-network_2020, scherbela2021solving, gao2021ab} and second-quantized \cite{choo2020fermionic} representations have recently been used to compute the ground-state energies of molecules to a level of accuracy rivaling, or in some cases exceeding, sophisticated quantum chemistry methods such as coupled cluster with singles, doubles, and perturbative triples \cite{shavitt2009}. The FermiNet and ans\"atze derived from it are the most accurate of these so far, gaining an advantage over second-quantized neural and most quantum chemical approaches because they are basis-set free. Choosing an appropriate basis set for a given system requires some understanding of the qualitative nature of the ground-state wave function. Freedom from this requirement, coupled with the flexibility of the neural representation, enables the application of the FermiNet to generic phases of matter.

We extend the FermiNet, which has previously only been applied to atoms and molecules \cite{pfau_ab_2020, spencer_better_2020, wilson2021simulations, li_fermionic_2022} \footnote{While we were preparing this manuscript, Wilson \emph{et al.\ }\cite{wilson2022wave} reported neural-network-based continuous-space VMC results for the same 14-electron system studied here, plus 7- and 19-electron gases. Although they used a more heavily modified version of the FermiNet ansatz, their results are similar to ours. They restrict their attention to the Fermi liquid phase and do not study the Wigner transition.}, to periodic systems. Recent work has used neural network ans\"atze to study periodic systems in continuous space, but has either focused on bosonic systems \cite{pescia_neural-network_2021} or used small basis sets \cite{yoshioka2021solving}, restricting their accuracy. 

We demonstrate the flexibility of the periodic FermiNet by studying the quantum phase transition between the Fermi liquid and Wigner crystal \cite{wigner_interaction_1934} in the three-dimensional interacting homogeneous electron gas (HEG) \cite{giuliani2005}. Two-dimensional Wigner crystals were very recently imaged for the first time \cite{zhou2021bilayer, smolenski2021signatures, li2021imaging}, but three-dimensional Wigner crystals have not yet been observed in electronic systems and are thus less well understood. The zero-temperature properties of the three-dimensional HEG depend on a single dimensionless parameter, $r_s$, defined as the ratio of the radius of a sphere that contains one electron on average to the Bohr radius. At high density (small $r_s$), the ground state is a weakly interacting Fermi liquid. At low density (large $r_s$), the correlations are stronger and the translational symmetry breaks spontaneously, giving rise to a spatially ordered  Wigner crystal \cite{wigner_interaction_1934}. 
We find that the same neural network architecture learns the appropriate ground-state wave function either side of the Wigner phase transition, spontaneously breaking continuous translational symmetry when the crystal phase is stable. As we give the network no information about the nature of the ground state, the degree of inductive bias in the determination is very low.

The Hamiltonian for a finite HEG of $N$ electrons subject to periodic boundary conditions is
\begin{equation}
    \label{eqn:hegham}
    \mathcal{H} = -\frac{1}{2}\sum_{i=1}^N\nabla^2_i + U_{\text{Coulomb}},
\end{equation}
where the indices $i$ label the $N$ electrons in the simulation cell and $U_{\text{Coulomb}}$ is the Coulomb energy per simulation cell of an infinite periodic lattice of identical copies of that cell \ In practice, the Coulomb energy is evaluated using the Ewald method \cite{ewald1921,fraser_finite-size_1996}. We work in Hartree atomic units, where energies are measured in Hartrees (1 Ha $\approx$ 27.211 eV) and distances in Bohr radii.

\begin{table*}[t]
    \begin{center}
\begin{tabular}{m{8em} P{8em} P{8em} P{8em} P{8em}} 
 \hline\hline
   & \multicolumn{4}{c}{Correlation energy [Hartree]} \\
  Method & $r_s=0.5$ & $r_s=1.0$ & $r_s=2.0$ & $r_s=5.0$ \\
  \hline
  SJB  & & & &\\
  \hspace{1em}VMC & $-0.58624(1)$ & $-0.5254(1)$ & $-0.437(3)$ & $-0.30339(2)$ \\
  \hspace{1em}DMC & $-0.58778(1)$ & $-0.5254(1)$ & $-0.4385(3)$ & $-0.30474(8)$ \\
 FermiNet & & & & \\
  \hspace{1em}$n_\text{det} = 1$ & $-0.58895(6)$ & $-0.52568(3)$ & $-0.43881(1)$ & $-0.30468(1)$ \\ 
  \hspace{1em}$n_\text{det} = 16$ & $-0.59094(6)$ & $-0.52682(3)$ & $-0.44053(1)$ & $-0.30495(1)$ \\
 i-FCIQMC\cite{shepherd_investigation_2012} & & & & \\
 $\quad$ finite basis & $-0.5939(4)$ & $-0.5305(5)$ & $-0.4430(7)$ & $-0.304(1)$ \\
 $\quad$ basis set limit & $\mathbf{-0.5969(3)}$ & $\mathbf{-0.5325(4)}$ & $\mathbf{-0.4447(4)}$ & $\mathbf{-0.306(1)}$ \\
 \hline\hline
\end{tabular}
\end{center}
    \caption{Correlation energy of the spin unpolarized $N=14$ HEG with simple cubic boundary conditions. The i-FCIQMC energies \cite{shepherd_investigation_2012} were calculated using a basis of 778 plane-wave orbitals for $r_s=5.0$ or 2378 plane waves otherwise, corresponding to Hilbert spaces of $10^{24}$ and $10^{31}$ Slater determinants, respectively. The extrapolation of the i-FCIQMC results to the complete basis set limit may yield correlation energies that are 1--2 mHa too negative \cite{neufeld_heg_cc_2017}. The Slater-Jastrow-backflow (SJB) VMC and DMC results were calculated using the \textsc{casino} program \cite{casino2010,casino2020} with a single-determinant SJB trial wave function optimized using variance minimization and then energy minimization. The DMC results were extrapolated to zero time step. The FermiNet results were obtained as explained in the text.}
    \label{table:14heg}
\end{table*}

The wave function represented by a FermiNet is a sum of determinants of many-electron (\emph{not} one-electron) functions \cite{pfau_ab_2020, spencer_better_2020}:
\begin{equation}
\label{eqn:ferminet}
    \Psi(\{\mathbf{x}_j\}) = \sum_k^{n_\text{det}} \det\left[ \psi^k_i(\mathbf{x}_j; \{\mathbf{x}_{/j}\}) \right],
\end{equation}
where $\mathbf{x} = ({\mathbf{r}, \alpha})$ labels the spatial and spin coordinates of an electron, and the set $\{\mathbf{x}_{/j}\}$ includes all-electron coordinates except $\mathbf{x}_j$. Multiplicative coefficients are not required as they can be absorbed into the determinants. The many-electron orbital $\psi^k_i(\mathbf{x}_j; \{\mathbf{x}_{/j}\})$ depends on the coordinates $\mathbf{x}_j$ of the $j$-th electron, and, in a permutation-invariant fashion, on the set of all other electron coordinates. The use of many-electron orbitals makes a FermiNet determinant much more flexible than a Slater determinant of one-electron orbitals, and it has been shown \cite{hutter2020representing} that a single determinant of this form can represent any antisymmetric function. The proof of this theorem depends upon the construction of discontinuous functions that cannot be represented in practice by a finite network of a reasonable size. Nevertheless, a linear combination of a small number of FermiNet determinants has a much greater representational capacity than a linear combination of an equal number of Slater determinants \cite{pfau_ab_2020}.

It is convenient to work with spin-assigned wave functions \cite{foulkes_quantum_2001}, replacing $\Psi(\mathbf{r}_1, \alpha_1; \ldots, \mathbf{r}_N, \alpha_N)$ with a function of position alone: $\Psi(\mathbf{r}_1, \ldots, \mathbf{r}_N) \triangleq \Psi(\mathbf{r}_1, \uparrow; \ldots \mathbf{r}_{N_{\uparrow}}, \uparrow; \mathbf{r}_{N_{\uparrow}+1}, \downarrow; \ldots, \mathbf{r}_{N_{\downarrow}},\downarrow)$, where $N = N_{\uparrow} + N_{\downarrow}$ is the number of electrons and $N_{\uparrow} - N_{\downarrow} = 2S_z$ is the spin polarization. The spin-assigned wave function is only antisymmetric on interchange of the position coordinates of electrons of the same spin, but assigning the spins has no effect on expectation values of spin-independent operators. Relabeling the electron positions according to the assigned spins, a FermiNet determinant becomes (in block-matrix form, determinant label $k$ dropped for clarity),
\begin{equation}
    \det[\bm{\psi}] =
    \begin{vmatrix}
    \psi_i^\uparrow(\mathbf{r}_j^\uparrow; \{\mathbf{r}^\uparrow_{/j}\}, \{\mathbf{r}^\downarrow_{/j}\}) &  \psi_i^\uparrow(\mathbf{r}_j^\downarrow; \{\mathbf{r}^\uparrow_{/j}\}, \{\mathbf{r}^\downarrow_{/j}\}) \left.\right. \\[8pt]
    \psi_i^\downarrow(\mathbf{r}_j^\uparrow; \{\mathbf{r}^\uparrow_{/j}\}, \{\mathbf{r}^\downarrow_{/j}\}) & 
    \psi_i^\downarrow(\mathbf{r}_j^\downarrow; \{\mathbf{r}^\uparrow_{/j}\}, \{\mathbf{r}^\downarrow_{/j}\})
    \end{vmatrix}.
\end{equation}
The original FermiNet architecture assumed that the determinant above was block diagonal. We have found that removing this constraint provides a small but noticeable variational improvement. Dense determinants are hence used unless otherwise stated. A comparison between results obtained using dense and block-diagonal determinants can be found in the Supplementary Material.

The FermiNet uses a neural network to approximate the many-electron orbitals appearing in the determinants \cite{pfau_ab_2020}.
The network consists of two parallel streams, for processing one-electron and two-electron information. The one-electron stream is constructed of repeating blocks, where each block contains a nonlinear layer and a permutation-equivariant function.
The two-electron stream is a comparatively small fully-connected feed-forward network. 
The outputs of the one- and two-electron streams at each layer are fed into the permutation-equivariant functions.
The multiple outputs of the one-electron stream are fed through a final linear layer to produce the required number of many-electron functions, $\{\phi_i^{k\alpha}\}$. This may be generalized to complex-valued functions by doubling the output dimension of the final linear layer and taking pairs of the resulting outputs to represent the real and imaginary components of $\phi_i^{k\alpha}$. All results were obtained with real wave functions unless otherwise noted.
Finally, the network outputs are multiplied by a parameterized multiplicative envelope, $\mathbf{f}$, to produce the many-electron orbitals $\psi_i^{k\alpha}(\mathbf{r}) = f_i^{k\alpha}(\mathbf{r}) \phi_i^{k\alpha}(\mathbf{r})$. 
The electron position vectors $\mathbf{r}_i$ and norms $\norm{\mathbf{r}_i}$, and the electron-electron separation vectors $(\mathbf{r}_i - \mathbf{r}_j)$ and norms $\norm{\mathbf{r}_i - \mathbf{r}_j}$, are supplied as inputs to the network, the output of which is the value of the many-electron wave function corresponding to those inputs. Full details of the network architecture are given in Ref.~\cite{pfau_ab_2020} and the Supplementary Material. 

To adapt the FermiNet architecture to periodic systems, it is sufficient to modify the input features to ensure that periodic boundary conditions are satisfied.
Periodic input features are most easily expressed in the basis $\{\mathbf{a}_1, \mathbf{a}_2, \mathbf{a}_3\}$ of primitive Bravais lattice vectors of the simulation cell. An arbitrary vector $\mathbf{r}$ is written as $s_1 \mathbf{a}_1 + s_2 \mathbf{a}_2 + s_3 \mathbf{a}_3$ and the periodic input features corresponding to $\bm{r}$ are obtained from the fractional coordinates $s_i$ via the component-wise transformation $s_i\to\left(\sin(2\pi s_i), \cos(2\pi s_i)\right)$.
A periodic analogue of the Euclidean norm may be defined in terms of fractional coordinates as
\begin{equation}
    \label{eqn:periodic_norm}
\begin{split}
    \Vert{s}\Vert^2_p = \sum_{ij}&[1-\cos(2\pi s_i)]S_{ij}[1-\cos(2\pi s_j)]\\
    & + \sin(2\pi s_i)S_{ij}\sin(2\pi s_j),
\end{split}
\end{equation}
where $S_{ij} = \mathbf{a}_i\cdot\mathbf{a}_j$ acts as a metric tensor in the fractional coordinate system. This definition of the norm is smooth, periodic with respect to the simulation cell, and proportional to the Euclidean norm as $\mathbf{s} \to 0$. Unlike the simpler norm introduced in \cite{pescia_neural-network_2021}, it retains these properties for non-cubic simulation cells.
These changes are sufficient to satisfy the periodic boundary conditions, but we have found that convergence speed and asymptotic convergence are improved by including an envelope of the form
\begin{equation}
    f_i^{k\alpha}(\mathbf{r}) = \sum_m \left[ \nu_{im}^{k\alpha} \cos(\mathbf{k}_m\cdot\mathbf{r}) + \mu_{im}^{k\alpha}\sin(\mathbf{k}_m\cdot\mathbf{r}) \right],
    \label{eqn:pbc_real_envelope}
\end{equation}
for real wave functions, or
\begin{equation}
    f_i^{k\alpha}(\mathbf{r}) = \sum_m \nu_{im}^{k\alpha} \exp(\mathrm{i}\mathbf{k}_m\cdot\mathbf{r}),
    \label{eqn:pbc_complex_envelope}
\end{equation}
for complex wave functions. The $\mathbf{k}_m$ are simulation-cell reciprocal lattice vectors up to the Fermi wavevector of the noninteracting electron gas, and $\nu_{im}^{k\alpha}, \mu_{im}^{k\alpha}$ are learnable parameters. Finally, when simulating the electron gas, the absence of nuclei (and hence electron-nuclear cusps) removes the need to include the norms of the electron positions as inputs.

\begin{figure*}[t]
    \centering
    \includegraphics[width=1.0\textwidth]{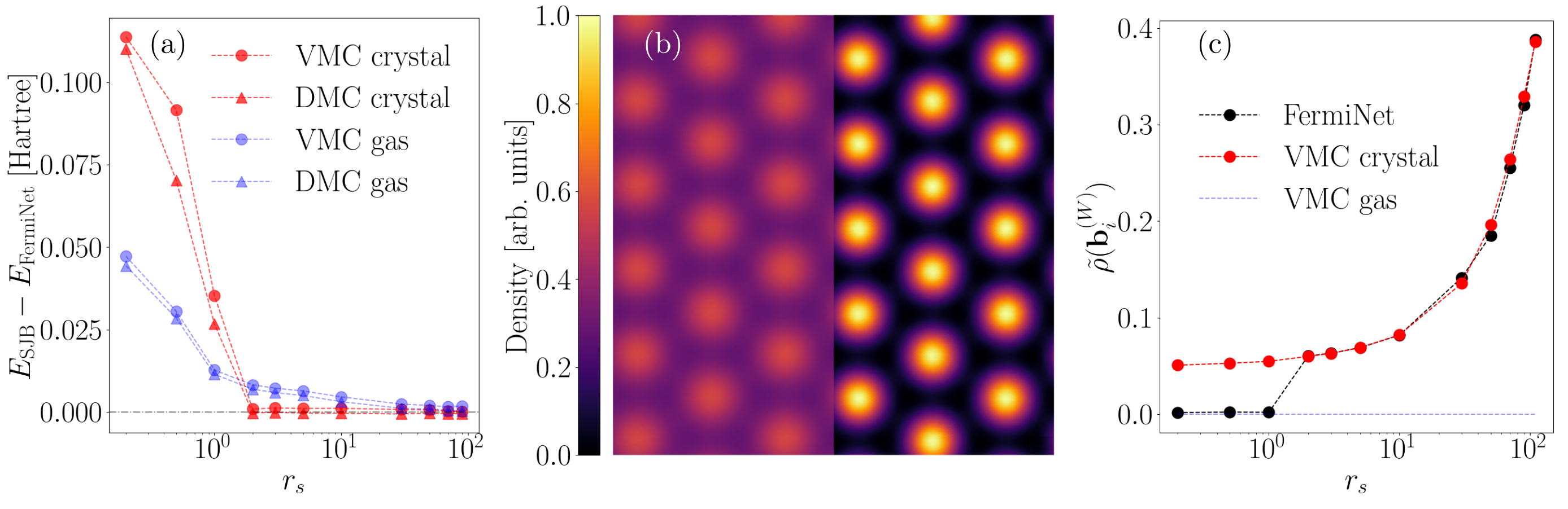} 
    \caption{\textbf{(a)} Single-determinant Slater-Jastrow-backflow (SJB) ground-state total energies per electron, measured relative to the FermiNet ground-state total energy per electron, of a spin-polarized 27-electron gas in a body-centered cubic (bcc) simulation cell. The ``gas'' and ``crystal'' results were obtained using SJB wave functions built using determinants of plane waves and Gaussian orbitals, respectively. Error bars are smaller than the markers. FermiNet results for $r_s\leq1$ used complex wave functions. FermiNet-VMC yields a variational improvement over SJB-VMC and SJB-DMC in the gas phase and over SJB-VMC in the crystal phase. The values used to construct this figure may be found in the Supplementary Material. \textbf{(b)} One-electron density of the $N=27$ spin-polarized HEG at $r_s=10$ (left) and $70$ (right), projected into the (011) plane of the conventional bcc structure, calculated via FermiNet-VMC\@. Four simulation cells are shown. Length scales are normalized by $r_s$ for comparison, such that the apparent length scales are equivalent and the crystal sites are superimposable. \textbf{(c)} Order parameter averaged over crystal axes for the bcc Wigner crystal state of the spin-polarized $N=27$ HEG\@. Error bars are smaller than the markers. At small values of $r_s$, the order parameter is $\sim$$0$, corresponding to a uniform one-electron density (gas-like); the order parameter rises sharply to a finite value at $r_s = 2$, corresponding to the emergence of a crystalline state.}
    \label{fig:27heg}
\end{figure*}

The FermiNet wave function is optimized using the variational Monte Carlo (VMC) method \cite{foulkes_quantum_2001}: the parameters of the network are adjusted to lower the expectation value of the energy, which is calculated using Metropolis-Hastings Monte Carlo integration over the $3N$-dimensional space of electron positions. Gradients of the energy are obtained using standard back-propagation techniques, and the network parameters are updated using the Kronecker-factored approximate curvature algorithm \cite{martens_optimizing_2020}, which approximates natural gradient descent \cite{amari_natural_1998} in a way that scales to large neural networks. Natural gradient descent is equivalent, up to a normalization constant, to the stochastic reconfiguration method \cite{sorella_wave_2005} frequently used in VMC \cite{nomura_rbm_2017,pfau_ab_2020}. Unless specified, all calculations used the same hyperparameters as in Ref.~\cite{pfau_ab_2020} which are given in the Supplementary Material.

\cref{table:14heg} shows the results of FermiNet calculations of the total energy of a 14-electron simple cubic simulation cell of unpolarized HEG at four different densities. This system is sufficiently small that near-exact initiator full configuration interaction quantum Monte Carlo (i-FCIQMC) benchmarks are available \cite{shepherd_investigation_2012}. 
The i-FCIQMC method \cite{cleland2011study} performs a stochastic diagonalization in a finite basis set, enabling the study of Hilbert spaces far larger than with exact diagonalization \cite{Booth2009-pu, cleland2011study}. However, the fermion sign problem in FCIQMC increases rapidly with $r_s$, rendering i-FCIQMC calculations at low densities with large basis sets impractical; the calculations at $r_s=5$ were ${\sim}10^4$ times more expensive than those at $r_s=1$ \cite{shepherd_investigation_2012}.
\cref{table:14heg} also includes VMC results calculated using a conventional Slater-Jastrow-backflow (SJB) wave function, and fixed-node diffusion Monte Carlo (DMC) results based on the VMC-optimized SJB wave function. The parameters of the VMC SJB wave function were optimized using variance minimization and then energy minimization, as implemented in the \textsc{casino} code \cite{casino2010, casino2020}. The DMC results were extrapolated to zero time step.

Although FermiNet is a VMC method, it achieves an accuracy similar to that of SJB-DMC, with both approaches obtaining 99\% of the i-FCIQMC correlation energy extrapolated to the complete basis set limit (which may be 1--2 mHa too large \cite{neufeld_heg_cc_2017}). FermiNet obtains a similar fraction of the correlation energy for molecular systems with a comparable number of electrons \cite{pfau_ab_2020}. Again as in molecular systems, calculations using sixteen FermiNet determinants are noticeably better than calculations using one FermiNet determinant.  

To assess the performance of FermiNet as the strength of the correlation increases, we study the $N$=$27$ electron spin-polarized HEG in the density range from $r_s=1$ to $90$. Prior work \cite{ceperley_ground_1980, drummond_diffusion_2004} had found Wigner crystallization to occur in the interval $r_s=[100, 110]$, although a recent study \cite{azadi_low-density_2022} lowers this estimate substantially. The 27-electron system studied here is very small and there are substantial finite-system-size effects that broaden the phase transition and move it to a much higher density.

Ground-state energies obtained using VMC with a FermiNet wave function and using VMC and DMC with SJB wave functions targeted at gas and crystal states are compared in \cref{fig:27heg}(a) for the 27-electron system. FermiNet energies for $r_s\leq1$ were obtained using complex wave functions, as discussed below. The precise form of the SJB ans\"{a}tze used to describe gases and crystals are detailed in the Supplementary Material and Ref.~\cite{drummond_diffusion_2004}. FermiNet VMC calculations produce a tighter variational lower bound than both the SJB gas and crystal wave functions at all densities. Furthermore, FermiNet outperforms fixed-node DMC calculations based on a SJB gas wave function across the entire density range, even at $r_s\leq1$. In the low-density regime, fixed-node DMC calculations using the SJB crystal wave function give slightly better results than our FermiNet VMC calculations. These results suggest that the nodal surface of the SJB crystal wave function is highly accurate but that the shape of the wave function away from the nodal surface is captured better by the FermiNet.

The Wigner crystal ground state of the HEG in the low-density limit is expected to be body-centered cubic (bcc) as this structure minimizes the packing density and has the lowest Madelung energy \cite{wigner_interaction_1934}.
The emergent localization of the wave function due to Wigner crystallization can be seen by accumulating the expectation value of the one-electron density operator,
\begin{equation}
    \label{eqn:density_operator}
    \rho(\mathbf{r}) = \left\langle\frac{1}{N}\sum_i\delta(\mathbf{r}_i - \mathbf{r})\right\rangle,
\end{equation}
where the expectation value is taken over samples of one-electron coordinates.
An order parameter for the broken-symmetry state is the Fourier component of $\rho(\mathbf{r})$ corresponding to any primitive reciprocal lattice vector, $\mathbf{b}_i^{\text{W}}$, of the emergent crystal:
\begin{equation}
    \label{eqn:orderparam}
    \Tilde{\rho}\bigl(\mathbf{b}_i^{\text{W}}\bigr) = \left\langle\frac{1}{N} \sum_j\text{exp}\left(i\mathbf{b}_i^{\text{W}}\cdot\mathbf{r}_j\right) \right\rangle .
\end{equation}
A state with  $\Tilde{\rho} = 0$ is gas-like, and a state with $\Tilde{\rho} \neq 0$ is crystalline. 
If the simulation cell is bcc in shape and contains $N = M^3$ (spin-polarized) electrons at low enough density, it contains an $M \times M \times M$ Wigner lattice and $\mathbf{b}_i^{\text{W}} = M\mathbf{b}_i$, where the $\mathbf{b}_i$ are the primitive reciprocal lattice vectors corresponding to the simulation cell.

Scans of the one-electron density corresponding to the optimized FermiNet wave functions at $r_{\text{s}} = 10$ and $70$ are shown in \cref{fig:27heg}(b). The figure shows the density in the $(\mathbf{a}_2, \mathbf{a}_3)$ plane, which is normal to the (011) direction of the conventional bcc cell. \cref{fig:27heg}(c) shows the order parameter $\Tilde{\rho}$ as calculated from VMC simulations using the FermiNet and SJB gas and crystal wave functions. These results show that FermiNet is capable of learning wave functions in both the gas and Wigner crystal states to very high accuracy without any hand-crafted features indicating whether the wave function should be localized or diffuse, any specific designation of crystal sites, or any other information that a transition should occur. We emphasize again that, unlike the gas and crystal SJB trial wave functions required to describe the gaseous and crystalline states accurately, the form of the FermiNet ansatz is identical across the entire density range.

For $r_s\leq1$, real-valued FermiNets often become trapped in local minima during optimization, with energies typically ${\sim}0.1$\% higher than the SJB-DMC benchmarks. However, complex-valued FermiNets do not become trapped in local minima at the same densities. The Hamiltonian here is real-valued, so the ground state can be taken to be a real function, and the converged complex FermiNets yield real-valued wave functions multiplied by a trivial (uniform in space) complex phase. This indicates that using a complex wave function improves optimization, rather than simply increasing representational capacity. Optimization also frequently becomes stranded in local minima for densities close to the phase transition in the crystalline regime ($r_s=2,3$, and $5$); however, this is reliably avoided by utilizing a higher learning rate, detailed in the Supplementary Material.

The center-of-mass coordinate of the electrons in the HEG separates and the wave function can be factored into a center-of-mass term, which is constant in the ground state, and a term that depends only on the vector separations of electrons. The one-electron density of the true ground state is thus uniform, not crystalline as we have found, and the crystalline order at low density appears in the pair-correlation function, not the one-electron density. This is known as a ``floating crystal'' state \cite{bishop1982, lewin2019}. As the size of the simulation cell tends to infinity, the cost of localizing the center of mass reduces to zero, the floating crystal becomes degenerate with the corresponding state of broken translation symmetry, and the phase transition is believed to become first order \cite{azadi_low-density_2022}, with the order parameter jumping from zero to a finite value at the critical density.


In Refs.\ \cite{drummond_diffusion_2004, azadi_low-density_2022}, it is shown (by considering a Slater determinant of Gaussian orbitals, with widths given by an empirical formula) that the energy difference between the fixed and floating crystal is approximately
\begin{equation}
    \Delta E = 0.055 r_s^{-3/2}.
\end{equation}
While the FermiNet differs from the Slater-type wave functions used to derive $\Delta E$, we expect a similar reduction in kinetic energy. At low $r_s$, $\Delta E$ is large ($20$mHa at $r_s=2$), so we would expect FermiNet to learn the floating crystal state. \cref{fig:27heg} (b, c) show that FermiNet instead learns the fixed crystal. The notion of a fixed origin can be removed by removing the one-electron features, however we find this increases the energy obtained. This suggests that the two-electron stream is insufficiently flexible to fully describe the two-electron correlations in the Wigner crystal without help from the one-electron stream. Improving the flexibility of the two-electron stream will be the focus of future work. We do not believe that these issues impact the central conclusion of the present work, and stress that in real condensed matter systems the Hamiltonian does not possess continuous translational symmetry.

To summarize, we have extended the FermiNet neural wave function to calculations with periodic boundary conditions. This we accomplished by making minimal, physically-motivated, modifications to render the input features periodic, and by adding a periodic envelope function. As proof of concept, we have demonstrated the accuracy of the modified architecture on the $N$=$14$ HEG, where we obtained ${\sim}99\%$ of the correlation energy and slightly outperformed VMC and DMC calculations using conventional one-determinant SJB trial wave functions. For the $N$=$27$ HEG, we see that the FermiNet is capable of learning the localized Wigner crystal phase \emph{a priori}, producing energies in excellent agreement with SJB trial wave functions which encode the qualitative nature of the ground state in their construction. This suggests that the FermiNet may be capable of determining novel quantum phases in condensed matter given only the Hamiltonian.

To study quantum phase transitions in realistic strongly-correlated electronic systems, it will be necessary to scale to larger numbers of electrons to overcome finite-size effects. This may require additional innovations in the neural network architecture employed. The FermiNet could also be used as a trial wave function for DMC calculations in periodic boundary conditions, an approach that yields small improvements in molecular systems \cite{wilson2021simulations}. More generally, we believe that the flexibility and accuracy offered by neural networks make them promising tools for studying complex correlation effects and other emergent phenomena.
The advantages of neural-network-based methods are most compelling when the phenomena in question are unexpected or not yet understood.
 
\begin{acknowledgments}
This work was undertaken with funding from the UK Engineering and Physical Sciences Research Council (EP/T51780X/1) (GC) and the Aker scholarship (HS)\@. Calculations were carried out with resources provided by the Baskerville Accelerated Compute Facility through a UK Research and Innovation Access to HPC grant. Via his membership of the UK's HEC Materials Chemistry Consortium, which is funded by EPSRC (EP/R029431), Foulkes used the UK Materials and Molecular Modelling Hub for computational resources, MMM Hub, which is partially funded by EPSRC (EP/T022213). Our SJB-DMC Wigner crystal calculations were performed using the Lancaster University's High-End Computing cluster, and the ARCHER2 UK National Supercomputing Service (http://www.archer2.ac.uk) via our membership of the UK's HEC Materials Chemistry Consortium, which is funded by EPSRC (EP/R029431).
\end{acknowledgments}

\bibliography{main}

\clearpage
\section{Supplementary}

\section{Fermionic neural networks}

The FermiNet architecture maps a set of input features derived from the electron coordinates, $\{\mathbf{r}_j^\alpha\}$, where $\alpha$ labels the spin of the electron, to the set of functions $\psi_i^{k\alpha}(\mathbf{r}_j; \{\mathbf{r}_{/j}\})$. In the original FermiNet, this set of inputs to each layer of the network is
\begin{equation}
\label{eqn:inputs}
    \mathbf{j}_i^{l\alpha} = \left(\mathbf{h}_i^\alpha, \frac{1}{n^\uparrow}\sum_{j=1}^{n^\uparrow}\mathbf{h}_j^{l\uparrow}, \frac{1}{n^\downarrow}\sum_{j=1}^{n^\downarrow}\mathbf{h}_j^{l\downarrow}\frac{1}{n^\uparrow}\sum_{j=1}^{n^\uparrow}\mathbf{h}_{ij}^{l\alpha\uparrow},  \frac{1}{n^\downarrow}\sum_{j=1}^{n^\downarrow}\mathbf{h}_{ij}^{l\alpha\uparrow} \right)
\end{equation}
where
\begin{eqnarray}
    \label{eqn:onefeatures}
    \mathbf{h}_i^{1\alpha} = \left(\mathbf{r}_i^\alpha - \mathbf{r}_I, \norm{\mathbf{r}_i^\alpha - \mathbf{r}_I} \quad\forall\quad I\right) \\
    \label{eqn:twofeatures}
    \mathbf{h}_{ij}^{1\alpha\beta} = \left(\mathbf{r}_i^\alpha - \mathbf{r}_j^\beta, \norm{\mathbf{r}_i^\alpha - \mathbf{r}_j^\beta}\right),
\end{eqnarray}
with capitalized subscripts referring to atomic co-ordinates, and $\norm{.}$ the Euclidean norm. These input features are updated by consecutive transformations,
\begin{eqnarray}
    \label{eqn:input_features_single}
    \mathbf{h}^{l+1\alpha}_i = \tanh\left(\underline{\underline{\mathbf{V}}}^l\mathbf{j}^{l\alpha}_i + \mathbf{b}^l\right) + \mathbf{h}^{l\alpha}_i \\
    \label{eqn:input_features_double}
    \mathbf{h}^{l+1\alpha\beta}_{ij} = \tanh\left(\underline{\underline{\mathbf{W}}}^l\mathbf{h}^{l\alpha\beta}_{ij} + \mathbf{c}^l\right) + \mathbf{h}^{l\alpha\beta}_{ij}.
\end{eqnarray}
The first transformation (one subscripted index) is referred to as the one-electron stream, and the second (two subscripted indices) the two-electron stream. The outputs from the $L$\textsuperscript{th} transformation are subject to a final, spin-dependent, linear transformation and multiplied by (in open boundary conditions) an exponentially decaying envelope \cite{spencer_better_2020} which enforces the decay of the wave function as $\mathbf{r}_i\to\infty$,
\begin{gather}
\label{eqn:orbitals}
\psi_i^{k\alpha}(\mathbf{r}_j, \{\mathbf{r}_{/j}\}) =\phi_i^{k\alpha}(\mathbf{r}_j, \{\mathbf{r}_{/j}\})f^{k\alpha}_i(\mathbf{r}_j)
\intertext{where}
\phi_i^{k\alpha}(\mathbf{r}_j, \{\mathbf{r}_{/j}\}) = (\mathbf{w}_i^{k\alpha}\cdot\mathbf{h}^{L\alpha}_j + g_i^{k\alpha})
\intertext{and}
f_i^{k\alpha}(\mathbf{r}_j) = \left[\sum_m\pi_{im}^{k\alpha}\text{exp}\left(-\sigma_{im}^{k\alpha}|(\mathbf{r}_j^\alpha - \mathbf{R}_m)|\right)\right].
\end{gather}
The functions $\psi_i^{k\alpha}$ are used as the inputs to the determinants, (\cref{eqn:ferminet}), in the main text. Note that the pooling operations in (\cref{eqn:inputs}) are chosen such that this feature vector is only permutation-invariant with respect to the exchange of electrons of the same spin. Thus, the desired fermionic exchange statistics are enforced even with a dense determinant. To construct complex wave functions, the number of functions $\phi_i^{k\alpha}$ output from the network is doubled, and the inputs to the determinant become
\begin{equation}
    \psi^{k\alpha}_i = (\phi_{2i}^{k\alpha} + \mathrm{i}\phi_{2i + 1}^{k\alpha})f_i^{k\alpha}
\end{equation}

The set of parameters,
\begin{equation}
\theta = \{\underline{\underline{\mathbf{V}}}^l, \underline{\underline{\mathbf{W}}}^l, \mathbf{w}_i^{k\alpha}, \mathbf{b}^l, \mathbf{c}^l, g_i^{k\alpha}, \pi_{im}^{k\alpha}, \sigma_{im}^{k\alpha}\},
\end{equation}
are all learnable. Pretraining these parameters to minimize the deviation between FermiNet orbitals and Hartree-Fock orbitals is possible, but we find that it is often unnecessary to achieve a well converged result and it is not performed in any calculations in the current study. The linear transformations specified by $\underline{\underline{\mathbf{V}}}^l$ and $\underline{\underline{\mathbf{W}}}^l$ are known as hidden layers. For a more extensive description of the FermiNet architecture, see Pfau \textit{et al.}\ \cite{pfau_ab_2020}.

FermiNets are trained via the variational Monte Carlo (VMC) method, a detailed description of which is provided by Foulkes \textit{et al.}\ \cite{foulkes_quantum_2001}. The parameters $\theta$ are optimized via gradient descent to minimize $\langle\mathcal{H}\rangle$. This guides the wave function $\Psi_\theta$ toward the ground state as a result of the variational principle. Working in log-space, the gradient of $\langle\mathcal{H}\rangle$ with respect to the parameters $\theta$ is
\begin{equation}
    \nabla_\theta\langle\mathcal{H}\rangle = \left\langle E_L\nabla_\theta\text{log}\Psi^* + E_L^*\nabla_\theta\text{log}\Psi - 2\left\langle E_L\right\rangle\nabla_\theta\text{log}|\Psi|\right\rangle,
\end{equation}
where $E_L(\mathbf{r})=\Psi^{-1}(\mathbf{r})\mathcal{H}\Psi(\mathbf{r})$ and the expectation value is evaluated for samples of $\mathbf{r}$ taken from the probability amplitude $|\Psi(\mathbf{r})|^2$. The kinetic component of the local energy is calculated in log-space via,
\begin{equation}
    T_L(\mathbf{r}) = -\frac{1}{2}\sum_i\left[\frac{\partial^2 \log\Psi}{\partial r_i^2}\Bigr|_\mathbf{r} + \left(\frac{\partial \log\Psi}{\partial r_i}\Bigr|_\mathbf{r}\right)^2\right].
\end{equation}

We employ the Kronecker-factored approximate curvature algorithm \cite{martens_optimizing_2020} which uses an approximation to the Fisher information matrix to carry out natural gradient descent \cite{amari_natural_1998}. For complex wave functions, the inverse Fisher matrix is not strictly the appropriate choice of metric for natural gradient descent, and one should instead use the Fubini-Study metric \cite{stokes_quantum_2020}. In the present work, we found that natural gradient using the usual inverse Fisher was capable of finding the ground state, possibly because the Hamiltonian is real-valued. A full treatment of the optimization of complex wave functions is beyond the scope of the present work.

\section{Periodic boundary conditions}
The ground state wave function of an interacting system possesses a macroscopically large number of degrees of freedom $n$, due to the many-body interactions between all of the charges in the system. Solving for the many-body wave function $\Psi(\mathbf{r}_1,\ldots,\mathbf{r}_n)$ in $\mathbb{R}^{3n}$ is intractable for $n$ approaching the---effectively infinite on a computational scale---number of electrons found in a real solid. 

To approximate real solids by simulating a small number of electrons we employ periodic boundary conditions: a finite-sized simulation cell is embedded in a periodic array of images of all charges in the simulation cell. The resulting Hamiltonian possesses discrete translational symmetry: displacing any charge by a simulation cell lattice vector leaves the system invariant. The many-body eigenfunctions then have the property \cite{rajagopal_variational_1995}
\begin{equation}
\label{eqn:supercell_bloch}
    \Psi(\mathbf{r}_1,\ldots,\mathbf{r}_i,\ldots,\mathbf{r}_n) = \Psi(\mathbf{r}_1,\ldots,\mathbf{r}_i+\mathbf{R}_\text{S},\ldots,\mathbf{r}_n),
\end{equation}
where $\mathbf{R}_\text{S}$ is a simulation cell lattice vector. As a result, the problem of finding eigenfunctions on $\mathbb{R}^{3n}$ for extremely large $n$ has been reduced to a problem of finding eigenfunctions on the torus  $\mathbb{T}^{3n}$ where $n$ is a small number. The errors arising due to this approximation are known as finite-size effects. A full treatment of finite-size effects are beyond the scope of the present work, and do not alter the conclusions of the comparisons presented as all systems being compared are utilizing the same finite-size Hamiltonian.

\section{FermiNet with periodic boundary conditions}

To impose the constraint (\cref{eqn:supercell_bloch}) on the FermiNet with it is sufficient to choose an alternative set of input features to the first layer of the FermiNet which are invariant under the translation of any one electron coordinate by a simulation cell lattice vector. In the following sections we describe modifications to the coordinate and distance features which fulfill this requirement.
\subsection{Fractional coordinates}
Any vector in real space can be expressed as a linear combination of primitive simulation cell lattice vectors $(\mathbf{a}_1, \mathbf{a}_2, \mathbf{a}_3)$,
\begin{equation}
    \mathbf{v} = s_1\mathbf{a}_1 + s_2\mathbf{a}_2 + s_3\mathbf{a}_3,\quad s_1, s_2, s_3 \in \mathbb{R},
\end{equation}
defining $\mathbf{s} = (s_1, s_2, s_3) \in \mathbb{R}^3$, which is an equally valid representation of a position on the lattice that we will refer to as fractional coordinates. There is a one-to-one mapping between positions in real space and positions in fractional coordinates,
\begin{equation}
    \mathbf{s} = \underline{\underline{\mathbf{A}}}^{-1}\mathbf{v}
\end{equation}
where,
\begin{equation}
    \underline{\underline{\mathbf{A}}} = \begin{pmatrix}
    \vert & \vert & \vert  \\
    \mathbf{a}_1 & \mathbf{a}_2 & \mathbf{a}_3 \\
    \vert  & \vert  & \vert 
    \end{pmatrix},
\end{equation}
is a matrix whose columns consist of the primitive simulation cell lattice vectors. The simulation cell is a parallelepiped in real space but a unit cube in fractional coordinates. As a result, it is simple to construct maps that are periodic under translations by simulation cell lattice vectors using trigonometric functions. 

All vectors $\mathbf{v}$ in the original set of input features [Eqs.\ (\ref{eqn:onefeatures}) and (\ref{eqn:twofeatures})] are replaced via the component-wise mapping
\begin{equation}
    \label{eqn:periodic_coords}
    v_n\to\left(\text{sin}(2\pi s_n), \text{cos}(2\pi s_n)\right).
\end{equation}
An additional subtlety is introduced by the fact that any \textit{continuous}, unique labeling of points on a unit circle requires two numbers. This necessitates the use of both sine and cosine input features for each spatial dimension by recognizing that the torus $\mathbb{T}^{3n}$ decomposes into a product of unit circles, $\mathbb{S}^n\times\mathbb{S}^n\times\mathbb{S}^n$. Displacing $\mathbf{v}$ by a simulation cell lattice vector leaves the value of the right-hand side of (\cref{eqn:periodic_coords}) invariant as desired.

\subsection{Periodic norm}

The periodic analogue of $||\mathbf{v}||$ must retain a cusp as $\mathbf{v}\to\mathbf{0}$, resembling the Euclidean norm. The inclusion of the Euclidean norm features is known to have a substantial impact on the accuracy of the FermiNet \cite{pfau_ab_2020}: the network is incapable of introducing discontinuities into the wave function, and thus cannot satisfy the Kato cusp conditions \cite{kato_eigenfunctions_1957} on the derivatives of the wave function as electrons approach nuclei and each other without cusps being included explicitly in the input features. Similarly, the periodic norm must be continuous everywhere except at the cusps, as the network will be unable to remove these discontinuities from the wave function, resulting in an unphysical contribution to the kinetic energy. In summary, we require a function of $\mathbf{s}$ which behaves like $|\mathbf{v}|=\sqrt{v_x^2+v_y^2+v_z^2}$ as $\mathbf{v}\to\mathbf{R}$, is periodic on the domain $[0,1]^3$, and whose derivative vanishes at the simulation cell boundaries to ensure continuity.

We proceed by considering the definition of the Euclidean norm as the Euclidean inner product of a vector with itself,
\begin{equation}
\label{eqn:euclid}
    \norm{\mathbf{v}}^2 = \mathbf{v}\cdot\mathbf{v} = (\underline{\underline{\mathbf{A}}} \mathbf{s})^T(\underline{\underline{\mathbf{A}}}\mathbf{s}) = \sum_{ij}s_iS_{ij}s_j,
\end{equation}
where
\begin{equation}
    S_{ij} = \mathbf{a}_i\cdot\mathbf{a}_j.
\end{equation}
We conjecture by analogy that the norm in terms of the periodic co-ordinates (\cref{eqn:periodic_coords}) should be
\begin{equation}
\begin{split}
    \norm{s}^2_p = \sum_{ij}&[1-\cos(2\pi s_i)]S_{ij}[1-\cos(2\pi s_j)]\\
    & + \sin(2\pi s_i)S_{ij}\sin(2\pi s_j),
\end{split}
\end{equation}
This definition of the norm possesses all of the properties that we desired: as $s_i\to0$ this expression reduces to the Euclidean norm [Eq.\ (\ref{eqn:euclid})] by considering the first-order Taylor expansions of sine and cosine; the periodicity is obvious as the expression is invariant to translations $s_i \to s_i \pm 1$; and, as a result of reducing to the Euclidean norm at the origin, this function retains the desired cusps, while also being differentiable in the rest of the unit cell. All distances, $||.||$, in the original set of input features [Eqs.\ (\ref{eqn:onefeatures}) and (\ref{eqn:twofeatures})] are replaced by the periodic norm. 

\subsection{Periodic multiplicative envelope}

\begin{figure}[t]
    \centering
    \includegraphics[width=0.5\textwidth]{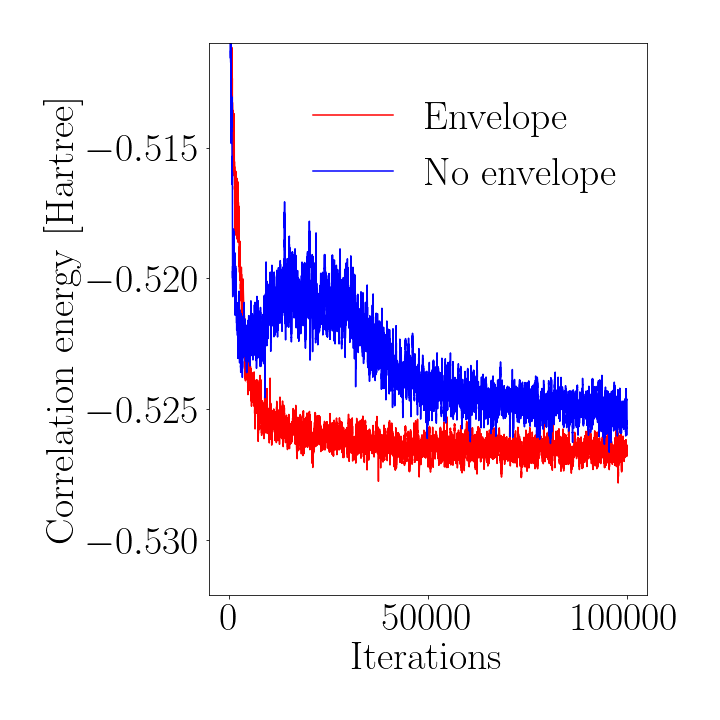}
    \caption{Learning curves, given in terms of the correlation energy, for the $N=14$ spin unpolarized HEG at $r_s = 1.0$ utilizing 16 dense determinants, with and without a sinusoidal envelope, using identical training parameters. An $N=500$ moving average filter has been applied to both curves to improve visual clarity.}
    \label{fig:envelope_lc}
\end{figure}

For the periodic envelope function introduced in \cref{eqn:pbc_real_envelope,eqn:pbc_complex_envelope}, $\nu_{im}^{k}$ and $\mu_{im}^{k}$ are strictly positive learnable parameters, optimized during training. The envelope eases the representation of highly oscillatory functions, while still being trivially capable of representing any function representable by the network with no envelope because $\mathbf{k}_0 = \mathbf{0}$. All $\nu_{im}^{k}$ and $\mu_{im}^{k}$ are initialized to small random values except $\nu_{i0}^{k}=1$. Other initialization schemes may be more appropriate, but have not been studied here. \cref{fig:envelope_lc} demonstrates the improved training performance due to the envelope.

\section{Experimental setup}

\subsection{FermiNet calculations}

\begin{table}
    \centering
    \small
    \begin{tabular}{|c|c|c|}\hline
        Kind & Parameter & Value \\\hline
       Optim & Batch size & 4096 \\
       Optim & Training iterations & 3e5 \\
       Optim & Pretraining iterations & 0 \\
       Optim & Learning rate & $(1e4 + t)^{-1}$ \\
       Optim & Local energy clipping  & 5.0 \\
       KFAC & Momentum & 0 \\
       KFAC & Covariance moving average decay & 0.95 \\
       KFAC & Norm constraint & 1e-3 \\
       KFAC & Damping & 1e-3 \\
       MCMC & Proposal std.\ dev.\ (per dimension) & 0.02 \\
       MCMC & Steps between parameter updates & 10 \\
       \hline
    \end{tabular}
    \caption{FermiNet hyperparameters for all experiments in the paper. These are mostly the same parameters as used in the original FermiNet paper, except we omit pretraining and slightly increase the number of training iterations.}
    \label{tab:hyperparams}
\end{table}

Four A100 GPUs were used for all calculations presented. Calculations were carried out using single precision floating point numbers, as we found statistically identical results were achieved using double precision at the cost of approximately doubling runtime. The modifications to FermiNet were implemented using the JAX Python library \cite{jax2018github}, extending a development version of the FermiNet \cite{ferminet_jax}. Optimization used a JAX implementation of the Kronecker-factored approximate curvature (KFAC) gradient descent algorithm \cite{spencer_better_2020, martens_optimizing_2020, kfac_jax}.
In the $N$=$14$ electron gas, a FermiNet with 4 layers of 256 units in the one-electron stream and 32 units in the two-electron stream was used for the 1 and 16 determinant calculations. In all calculations for the $N$=$27$ HEG, we used a FermiNet of four layers with 512/64 units in the one/two-electron streams respectively with 16 determinants. For both systems, the wave function was optimized over 3e5 training iterations and 5e4 additional samples of $\langle\mathcal{H}\rangle$ with the wave function parameters frozen were taken to obtain the final energies. The standard error associated with these energies was evaluated using a reblocking method \cite{flyvbjerg_error_1989} to account for sequential correlations introduced by the Monte Carlo sampling strategy.

For the majority of calculations, we employ a set of network and training hyperparameters identical to those previously used to obtain results on molecular systems \cite{pfau_ab_2020}. For the $N$=$27$ HEG at $r_s=2,3$, and $5$, we find that an initially more aggressive learning rate (base value 1e-2 versus the default 1e-4) is required to reliably avoid local minima. This is accompanied by an increased KFAC norm constraint and damping of 1.0 and 1e-1, respectively.

\subsection{Slater-Jastrow-backflow calculations}

The Slater-Jastrow-backflow (SJB) wave function ansatz used to produce the benchmark VMC and DMC calculations in the main text takes the form
\begin{equation}
    \Psi(\mathbf{r}) = \exp(J(\mathbf{r}))S(\mathbf{x}(\mathbf{r})).
\end{equation}
The Slater determinant $S$ is composed of one-electron functions and enforces the fermionic antisymmetry of the wave function, just as in the FermiNet. This determinant is evaluated at coordinates which are modified by a backflow transformation, 
\begin{equation}
    \mathbf{x}(\mathbf{r}) = \mathbf{r} + \bm{\xi}(\mathbf{r}),
\end{equation} 
and multiplied by a Jastrow factor $\exp(J)$ which is a permutation-invariant function of the electronic coordinates. Here we will only provide a brief overview of the terms incorporated into these factors. A much more detailed account is provided in Ref.\ \cite{azadi_low-density_2022}. All SJB VMC and DMC calculations were performed using the \textsc{casino} program \cite{casino2010,casino2020}

For the Fermi fluid, the one-electron orbitals in the Slater determinant are the Hartree-Fock orbitals for the homogeneous electron gas,
\begin{equation}
    \phi_\mathbf{k}(\mathbf{r}_j) = \exp(i\mathbf{k} \cdot\mathbf{r}_j),
\end{equation}
where the $\{\mathbf{k}\}$ are the $N/2$ (spin unpolarized) or $N$ (spin polarized) smallest simulation cell reciprocal lattice vectors. In the crystal, periodic one-electron orbitals are evaluated as sums over periodic images of site-centered Gaussian functions,
\begin{equation} 
\phi_{{\bf R}_\text{P}}({\bf r}_j) = \sum_{{\bf R}_\text{S}}
\exp\left(-C\left|{\bf r}_j-{\bf R}_\text{P}-{\bf
  R}_\text{S}\right|^2\right),
\end{equation}
where $\mathbf{R}_\text{P}$ is a primitive-cell lattice point within the simulation cell, $\mathbf{R}_\text{S}$ is a simulation cell lattice point, and $C$ is an optimizable parameter controlling the width of the Gaussian. This sum is truncated when the contributions of the images of the Gaussian basis functions become smaller than $10^{-7}$ at the edge of the simulation cell in which the orbital is being evaluated.  There are $N$ primitive-cell lattice points within the simulation cell.

The Jastrow exponent consists of a sum of three terms,
\begin{equation}
    J(\mathbf{r}) = \sum_{i<j}^N\left[u(r_{ij}) + p(\mathbf{r}_{ij})\right] + \sum_i^N q(\mathbf{r}_i),
\end{equation}
where $u$ is a power series in the electronic separations which includes fixed terms to impose the Kato cusp conditions \cite{kato_eigenfunctions_1957}. This term is smoothly cut off at a radius less than or equal to the radius of the largest sphere that can be inscribed in the Wigner-Seitz cell of the simulation cell. The $p$ term is 
\begin{equation}
 p(\mathbf{r}_{ij})=\sum_A a_A \sum_{\mathbf{G} \in A^+} \cos(\mathbf{G}
 \cdot \mathbf{r}_{ij}), \label{eq:p_term}
\end{equation}
where $A$ consists of shells of simulation cell reciprocal lattice vectors, and $A^+$ excludes one from each pair of vectors which are related by inversion symmetry. Similarly, \begin{equation}
  q(\mathbf{r}_i)=\sum_B b_B \sum_{\mathbf{G} \in B^+} \cos(\mathbf{G}
 \cdot \mathbf{r}_i),
\end{equation}
where $B$ consists of shells of Wigner crystal primitive cell reciprocal lattice vectors. The $q$ term is omitted from the Fermi fluid wave function, as it does not retain continuous translational invariance with respect to the electronic center of mass.

The backflow transformation consists of two terms,
\begin{equation}
    {\bm \xi}_i(\mathbf{r})=\sum_{j\neq i}^{N} \eta(r_{ij})
 \mathbf{r}_{ij} + \sum_{j\neq i}^N {\bm \pi}({\bf r}_{ij}),
\end{equation}
where $\eta$ is mathematically identical to the Jastrow $u$ term, and ${\bm \pi}$ has the form of the gradient of the Jastrow $p$ term:
\begin{equation}
    {\bm \pi}({\bf r}_{ij}) = -\sum_A c_A \sum_{{\bf G} \in A^+} \sin({\bf G}
\cdot {\bf r}_{ij}) \, {\bf G}.
\end{equation}
The coefficients $a_A$, $b_B$, and $c_A$ are all optimizable parameters.

All of these terms are evaluated using a minimum image convention.

\section{Block determinants}

In the main text we introduce the concept of dense determinants in the FermiNet architecture. Here we provide a comparison with energies for the unpolarized $N$=$14$ HEG, obtained using block diagonal determinants. These values are presented in \cref{table:14heg_blockdet}, alongside the results obtained using dense determinants, copied from the main text for comparison. In all cases dense determinants offer a small variational improvement in the correlation energy obtained.

\begin{table}[t]
    \begin{center}
\begin{tabular}{m{2.5em} P{5.5em} P{5.5em} P{5.5em} P{5.5em}} 
 \hline\hline
   & \multicolumn{4}{c}{Correlation energy [Ha]} \\
  $n_\text{det}$ &$r_s=0.5$ & $r_s=1.0$ & $r_s=2.0$ & $r_s=5.0$ \\
  \hline \\
 & \multicolumn{4}{c}{Block determinants} \\
1 &    $-0.58831(7)$ & $-0.52510(3)$ & $-0.43842(1)$ & $-0.30433(1)$\\ 
16 &   $-0.58962(6)$ & $-0.52558(3)$ & $-0.43876(1)$ & $-0.30466(1)$ \\
 & \multicolumn{4}{c}{Dense determinants} \\
1 & $-0.58895(6)$ & $-0.52568(3)$ & $-0.43881(1)$ & $-0.30468(1)$ \\ 
16 & $\mathbf{-0.59094(6)}$ & $\mathbf{-0.52682(3)}$ & $\mathbf{-0.44053(1)}$ & $\mathbf{-0.30495(1)}$ \\
 \hline\hline
\end{tabular}
\end{center}
    \caption{Correlation energy of the spin unpolarized $N=14$ HEG with simple cubic boundary conditions, compared between a FermiNet wave function constructed from block diagonal and dense determinants.}
    \label{table:14heg_blockdet}
\end{table}

\begin{table*}[p]
\caption{Total energy per electron of the $N$=$27$ HEG in a body-centered cubic simulation cell at a range of densities, obtained via variational Monte Carlo. SJB values were obtained from our calculations using the \textsc{casino} package. Here, ``gas'' refers to a Slater determinant of plane-wave orbitals, and ``crystal" refers to a Slater determinant of Gaussian orbitals. FermiNet results for $r_s\leq1$ used complex wave functions.}

\begin{center}
    \begin{tabular}{m{6em} P{9em} P{9em} P{9em} P{9em} P{9em}}
    \hline\hline
    & \multicolumn{5}{c}{Total energy [Hartree per electron]} \\
    $r_s$ & FermiNet & SJB-VMC Gas & SJB-DMC Gas & SJB-VMC Crystal & SJB-DMC Crystal \\
    \hline
    0.2 & $44.18067(2)$ & $44.18235(4)$ & $44.18224(1)$ & $44.184815(4)$ & $44.18468(3)$\\
0.5 & $6.318704(2)$ & $6.31983(2)$ & $6.319754(7)$ & $6.322090(8)$ & $6.32130(6)$\\
1.0 & $1.261547(1)$ & $1.262022(8)$ & $1.261975(2)$ & $1.26286(1)$ & $1.26254(2)$\\
2.0 & $0.152086(1)$ & $0.152393(3)$ & $0.152344(3)$ & $0.1521205(4)$ & $0.152077(4)$\\
3.0 & $-0.006961(2)$ & $-0.006693(2)$ & $-0.0067419(7)$ & $-0.006915(2)$ & $-0.006966(2)$\\
5.0 &  $-0.057691(4)$ & $-0.057453(1)$ & $-0.0575023(4)$ & $-0.0576496(9)$ & $-0.057702(1)$\\
10.0 & $-0.050320(5)$ & $-0.0501501(6)$ & $-0.0502038(1)$ & $-0.0502782(7)$ & $-0.0503349(8)$\\
30.0 & $-0.022541(5)$ & $-0.0224510(2)$ & $-0.02249870(5)$ & $-0.0225119(3)$ & $-0.0225620(2)$\\
50.0 & $-0.014478(4)$ & $-0.0144033(1)$ & $-0.01444409(5)$ & $-0.0144528(1)$ & $-0.0144918(1)$\\
70.0 & $-0.010693(3)$ & $-0.0106355(1)$ & $-0.0106735(1)$ & $-0.0106824(1)$ & $-0.0107118(1)$\\
90.0 & $-0.008496(1)$ & $-0.0084277(1)$ & $-0.0084862(1)$ & $-0.00849274(7)$ & $-0.0085137(2)$\\

    \hline\hline
    \end{tabular}
\end{center}

\label{table:27heg}
\end{table*}
\end{document}